\documentclass[twocolumn,prb,showpacs]{revtex4}
\usepackage{graphicx}
\usepackage{dcolumn}
\usepackage{bm}
\begin{document}
\title{Magnetic impurity in correlated electrons system}
\author{S.~Nishimoto and P.~Fulde}
\affiliation{Max-Planck-Institut f\"ur Physik komplexer Systeme, D-01187 Dresden, Germany}
\date{\today}
\begin{abstract}

We study a magnetic impurity embedded in a correlated electron system using the density-matrix 
renormalization group method. The correlated electron system is described by the one-dimensional 
Hubbard model. At half filling, we confirm that the binding energy of the singlet bound state 
increases exponentially in the weak-coupling regime and decreases inversely proportional to 
the correlation in the strong-coupling regime. The spin-spin correlation shows an exponential 
decay with distance from the impurity site. The correlation length becomes smaller with 
increasing the correlation strength. We find discontinuous reduction of the binding energy and 
of spin-spin correlations with hole doping. The binding energy is reduced by hole doping; 
however, it remains of the same order of magnitude as for the half-filled case.
\end{abstract}
\maketitle

\section{INTRODUCTION}

Although more than forty years have passed since the discovery of the Kondo effect, 
it is still one of the most interesting topics in condensed matter physics; it lies 
at the heart of understanding strongly correlated electron systems.~\cite{hewson} 
The Kondo effect, which leads to the quenching of an impurity spin, forms the basis of 
the physics of a single magnetic impurity embedded in a metal. In order to understand 
the Kondo effect, the Anderson model~\cite{anderson} has been applied with great success. 
In the theoretical studies, one generally assumes an impurity level to be embedded in a 
noninteracting conduction band.

In the past a system of magnetic ions coupled to (strongly) correlated conduction electrons 
has attracted considerable interest in connection with heavy-fermion behavior, namely  
Nd$_{2-x}$Ce$_x$CuO$_4$.~\cite{brugger} This raised the question whether correlations 
among conduction electrons affect substantially the expected formation of heavy 
quasiparticles.~\cite{fulde} So far a number of authors have studied models of a single 
magnetic impurity embedded in a host of correlated conduction electrons. 
Thereby perturbation theory and other approximation schemes were 
applied.~\cite{schork,poilblanc1,igarashi,khaliullin,hofstetter} For example, it was shown that 
the Kondo scale can increase exponentially in the weak-coupling regime with increasing 
interaction of the conduction electrons.~\cite{khaliullin,hofstetter} But a quantitative 
theory is still missing. Moreover the case of strongly correlated conduction electrons 
with band filling slightly less than one-half (hole doping) is still an open problem.

In this paper we study a single magnetic impurity coupled to a correlated electron system. 
The latter is assumed to be one dimensional (1D) and described by 
a Hubbard Hamiltonian. Using the density-matrix renormalization group (DMRG) method, 
we calculate the binding energy of the impurity-induced bound state and spin-spin 
correlation functions between the impurity and the correlated electrons in the thermodynamic 
limit. Special attention is paid to the case of a nearly half-filled conduction band with 
repulsive electron-electron interactions. For a 1D correlated host, there have been numerical 
study for similar models~\cite{hallberg} as well as an analytical study for a integrable 
model.~\cite{zvyagin} We hope that the present investigation will contribute to better insights.

This paper is organized as follows. In Sec.~II, we introduce our model, i.e., a magnetic 
impurity coupled to a Hubbard chain. In Sec.~III, we give some numerical details of the DMRG 
method applied here. In Sec.~IV, we first present calculated results for the binding energy 
and spin-spin correlation functions at half filling, and discuss the effect of the 
host-band correlations on the Kondo physics. Then, we consider the evolution of the same 
quantities with hole doping. Sec.~IV contains a summary of the results and the discussions.

\section{MODEL}

\begin{figure}[t]
    \includegraphics[width= 8.0cm]{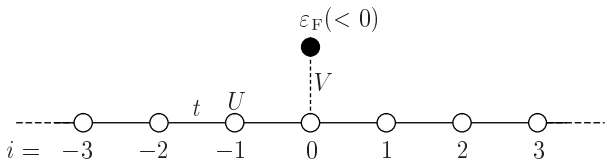}
  \caption{
Lattice structure of the system. Open and solid circles represent correlated electron 
system and impurity site, respectively. The bottom numbers $i$ denote site index of 
correlated electron system and $|i|$ corresponds to a distance between site $i$ and 
the impurity site. 
  }
    \label{fig1}
\end{figure}

We study a magnetic impurity coupled to a 1D correlated electron system. The Hamiltonian 
consists of three terms
\begin{eqnarray}
H = H_c + H_f + H_{cf}.
\label{hamiltonian}
\end{eqnarray}
The first term $H_c$ represents 1D correlated electrons. Here we describe them by a 
Hubbard Hamiltonian,
\begin{eqnarray}
H_c = t \sum_{i,\sigma}(c^\dagger_{i+1 \sigma}c_{i \sigma} + h.c.) 
+ U \sum_i n_{i \uparrow}n_{i \downarrow},
\label{hamconduction}
\end{eqnarray}
where $c_{i \sigma}^\dagger$ ($c_{i \sigma}$) is the creation (annihilation) operator of 
an electron with spin $\sigma$ ($=\uparrow, \downarrow$) at site $i$, and 
$n_{i\sigma}=c_{i \sigma}^\dagger c_{i \sigma}$ is the number operator. Furthermore $t$ is 
the hopping integral between neighboring sites and $U$ is the onsite Coulomb interaction. 
The second term $H_f$ is the orbital energy of the magnetic impurity site. We assume 
that the impurity contains one orbital, e.g., 4{\it f} and the Coulomb repulsion on the 
orbital $U_f$ is infinite. Since double occupancies are excluded, i.e., the {\it f} orbital 
is either empty or singly occupied, the impurity site is given by
\begin{eqnarray}
H_f = \varepsilon_f \sum_{\sigma} \hat{f}^\dagger_\sigma \hat{f}_\sigma,
\label{hamimpurity}
\end{eqnarray}
with $\hat{f}_\sigma^\dagger=f_\sigma^\dagger(1-f_{\bar{\sigma}}^\dagger f_{\bar{\sigma}})$ and 
$\varepsilon_f<0$. For convenience, we define $r=-\varepsilon_f/U$($>0$) and label an 
electron on the impurity site as ``{\it f} electron''. The third term $H_{cf}$ 
involves the interaction between the impurity site and the correlated electron system. 
The interaction is assumed to be local and described by a hybridization like in the Anderson 
model, i.e., the impurity site is hybridized with a single site (denoted as site $0$) of the 
correlated electron system. Thus, 
\begin{eqnarray}
H_{cf} = V \sum_{\sigma} (c_{0\sigma}^\dagger \hat{f}_\sigma 
+ \hat{f}^\dagger_\sigma c_{0\sigma}),
\label{hamhybridization}
\end{eqnarray}
where $V$ is the hopping integral between the impurity site and site $0$. The lattice structure 
is shown in Fig.~\ref{fig1}. We will work in units where $t=1$ and take as typical values 
$|\varepsilon_f|= 2-3$ and $V=0.1-0.2$, throughout.

\section{METHOD}

We employ the DMRG method, which is one of the most powerful numerical techniques for 
studying quantum lattice many-body systems including quantum impurity systems.~\cite{dmrg} 
With the DMRG method, we can obtain ground-state and low-lying excited-states energies, 
as well as expectation values of physical quantities quite accurately for very large 
finite-size systems. 

In order to carry out our calculations, we consider $N$ ($=N_\uparrow+N_\downarrow$) electrons 
($N$: even) in a system consisting of a chain of $L$sites correlated electron system ($L$: odd) 
and a single impurity site. The electron density is defined as $n=N/(L+1)$. Note that the number of 
lattice sites must be taken as $L+1=4l-2$ with $l$ ($>1$) being an integer to maintain the total 
spin of the ground state as $S=0$. If one chooses it as $L+1=4l$, the singlet and triplet states 
are degenerate. We now apply the open-end boundary conditions to the 1D correlated 
electron system and assume that the impurity site is hybridized with the central site of the 
1D open chain. The latter corresponds to site $0$, and sites $i$ and $-i$ are equivalent. In this paper we restrict ourselves to the half-filled and hole-doped cases ($N \le L+1$). 

Regarding quantum impurity problems, it is generally complicated for finite-size calculations 
to obtain accurate results in the thermodynamic limit $L \to \infty$ because of finite-size 
effects. In our calculations, the most problematic finite-size effects are Friedel oscillations 
due to the open ends of the Hubbard chain. Mostly, the energy scale of the Kondo physics is 
exponentially small; nevertheless, Friedel oscillations can persist even at the center 
of the chain as they decay as a power-law from the edge sites. Therefore, we study several 
long chains with sites $L+1=62$, $126$, $190$, $254$, $318$, $382$, $446$, and $510$, and 
then perform the finite-size-scaling analysis based on the size-dependent quantities. All DMRG 
results in this paper are extrapolated to the thermodynamic limit $L \to \infty$. For precise 
calculations we keep up to $m \approx 5000$ density-matrix eigenstates in the DMRG procedure. 
In this way, the maximum error in the ground-state energy is below $10^{-8}-10^{-7}$. 

\section{RESULTS}

\subsection{System at half filling ($n=1$)}

\subsubsection{Binding energy}

We first study the binding energy between the {\it f} electron and the correlated electrons. 
It corresponds to an energy gain due to the formation of a Kondo (or local) singlet bound 
state. Hence, the binding energy is given by an energy difference between the first triplet 
excited state and the singlet ground state,
\begin{eqnarray}
\Delta_{\rm B}=\lim_{L \to \infty} \Delta_{\rm B}(L),
\label{binddef1}
\end{eqnarray}
with
\begin{eqnarray}
\Delta_{\rm B}(L)=E_0(L,N_\uparrow+1,N_\downarrow-1)-E_0(L,N_\uparrow,N_\downarrow)
\label{binddef2}
\end{eqnarray}
where $E_0(L,N_\uparrow,N_\downarrow)$ is the ground state energy in a system of $L+1$ sites 
with $N_\uparrow$ up-spin and $N_\downarrow$ down-spin electrons. Note that, at half filling, 
the system is insulating for finite $U$. The bound state therefore may be from as 
a local singlet rather than the Kondo singlet. Here and in the following we will speak of a 
Kondo singlet only if it involves more than the central site of the correlated electrons.

\begin{figure}[t]
    \includegraphics[width= 7.5cm]{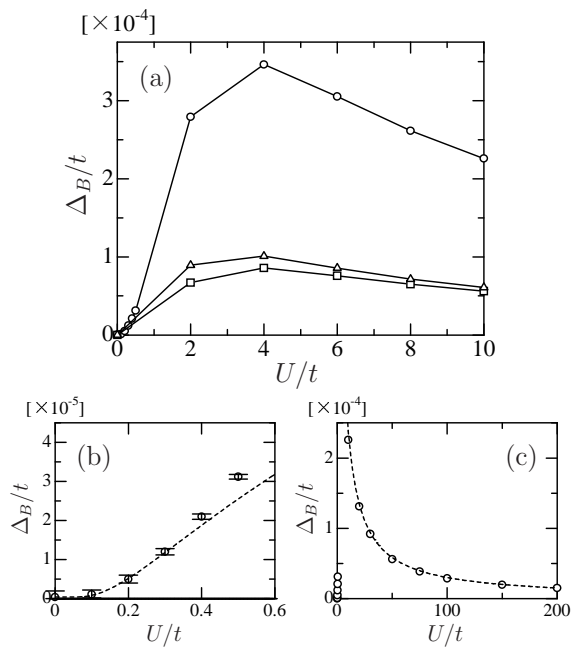}
  \caption{
(a) Binding energy $\Delta_{\rm B}$ for $\varepsilon_f=-3$, $V=0.2$ (circles), 
$\varepsilon_f=-3$, $V=0.1$ (triangles), and $\varepsilon_f=-2$, $V=0.1$ (squares). 
(b) Magnified view of small $U$ region for $\varepsilon_f=-3$, $V=0.2$. The data is 
fitted by a function $\Delta_{\rm B}=\sqrt{\alpha U} \exp(-\beta/U)$ 
with $\alpha \simeq 2.5 \times 10^{-4}$ and $\beta \simeq 0.4$. 
(c) $\Delta_{\rm B}$ for $\varepsilon_f=-3$, $V=0.2$ in the strong-coupling regime 
($U \gg t$). The data is fitted by a function 
$\Delta_{\rm B} \simeq \frac{\gamma}{U-\varepsilon_{\rm F}}$ with 
$\gamma \simeq 3.1 \times 10^{-4}$.
  }
    \label{fig2}
\end{figure}

In Fig.~\ref{fig2}(a), we show the DMRG results of the binding energy $\Delta_{\rm B}$ as 
a function of the Coulomb interaction $U$ for various parameter sets. In total, the results 
for the different parameter sets are qualitatively the same; as $U$ increases, the binding 
energy rises rapidly for small $U$, reaches a maximum around $U \approx 4$ and decreases 
gradually for large $U$. This behavior is similar to the dependence of the effective 
Heisenberg interaction on the Coulomb interaction in the half-filled Hubbard 
model.~\cite{szczech} Accordingly, the DMRG results show that for large values of $U$ 
the binding energy is 
approximately proportional to the effective exchange coupling $J_{cf}$, between the impurity 
and site $0$.~\cite{yosida} If we assume that the effective exchange coupling results from 
second-order perturbation, i.e., $J_{cf} = \frac{2V^2}{(U-\varepsilon_f)}$, 
we can explain why the results for $V=0.2$ are about four times larger than those for $V=0.1$. 
This estimation of the effective exchange coupling is also consistent with a slight decrease 
of the binding energy with increasing $|\varepsilon_f|$.

Let us now consider the behavior in the limiting cases for weak and strong interaction strengths. 
A magnified view of the weak-coupling regime ($U<t$) for $\varepsilon_f=-3, V=0.2$ is given in 
Fig.~\ref{fig2}(b). When $U=0$, the system is metallic and essentially equivalent to the 
single-impurity Anderson model (SIAM) in the Kondo limit ($U_f/V=\infty$) but asymmetric 
case ($\varepsilon \neq -U_f/2$). The orbital energy of the impurity site is lower than the 
Fermi energy of the conduction band, so that the occupation number of the impurity site is 
always $1$. The exchange interaction $J_{cf}$ is estimated to be the order of 
$V^2/\varepsilon_{\rm F}$ and, therefore, the binding energy is expected to be very small 
but finite. We estimate it to be roughly $\Delta_{\rm B} \simeq 10^{-7}-10^{-6}$. 
This value is compatible with the Kondo temperature $T_{\rm K}$ in the 
asymmetric SIAM.~\cite{meyer} The introduction of a finite Coulomb interaction makes 
the system insulating. With increasing $U$, $\Delta_{\rm B}$ increases gradually when 
$U/t \lesssim 0.2$ and rapidly for $U/t \gtrsim 0.2$. There is a crossover from the Kondo 
singlet to a local singlet around $U/t = 0.2$. Assuming an exponential behavior of 
$\Delta_{\rm B}$ with $U$ leads to a good fitting of the DMRG data, 
i.e., $\Delta_{\rm B}=\sqrt{\alpha U} \exp(-\beta/U)$ with $\alpha \simeq 2.5 \times 10^{-4}$ 
and $\beta \simeq 0.4$. Furthermore, $\Delta_{\rm B}$ increases almost linearly in the regime 
$U/t =0.2-2$. We thus find that the binding energy of the local singlet can be a few orders of 
magnitude larger than that of the Kondo singlet. 

The DMRG results for the strong-coupling regime ($U \gg t$) with $\varepsilon_f=-3$ and $V=0.2$ 
are plotted in Fig.~\ref{fig2}(c). In this regime, the electrons are strongly localized at 
each site. Therefore, the system (\ref{hamiltonian}) can be reduced to the Heisenberg model 
with Hamiltonian
\begin{eqnarray}
H_{\rm eff} = J \sum_i {\bf s}_i \cdot {\bf s}_{i+1} + J_{cf} {\bf S}_f \cdot {\bf s}_0
\label{Heisham}
\end{eqnarray}
with $J=\frac{4t^2}{U}$. The DMRG data can 
be fitted quite well by a function $\Delta_{\rm B} = \frac{\gamma}{U-\varepsilon_{\rm F}}$ 
with $\gamma \simeq 3.1 \times 10^{-4}$. Despite the strong localization of the electrons 
the binding energy is two orders of magnitude smaller than the {\it cf} exchange coupling. 
This is because for $n=1$ a spin-density-wave (SDW) is forming in the chain for any value of 
$U$ ($>0$) which makes the formation of the local singlet state more difficult. We note that 
the behavior of the binding energy for finite $U$ is essentially the same as that of 
the N{\'e}el temperature in the half-filled Hubbard model.~\cite{szczech}

\subsubsection{Spin-spin correlations}

\begin{figure}[t]
    \includegraphics[width= 6.5cm]{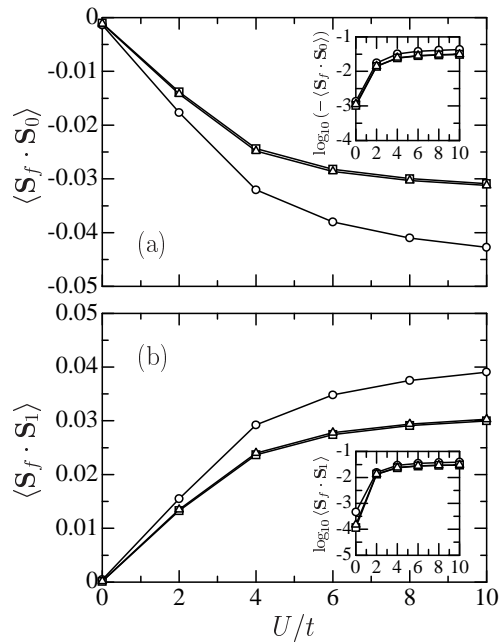}
  \caption{
Spin-spin correlation functions (a) $\left\langle {\bf S}_f \cdot {\bf s}_0 \right\rangle$ and 
(b) $\left\langle {\bf S}_f \cdot {\bf s}_1 \right\rangle$ as a function of  
the Coulomb interaction $U$ for $\varepsilon_f=-3$, $V=0.2$ (circles), 
$\varepsilon_f=-3$, $V=0.1$ (triangles), and $\varepsilon_f=-2$, $V=0.1$ (squares). 
Inset: semilogarithmic plots of the magnitude of the spin-spin correlation functions.
  }
    \label{fig3}
\end{figure}

In the Kondo problem, the spin degrees of freedom around the impurity play 
an essential role. Therefore, we investigate spin-spin correlations between the {\it f} 
electron and the correlated electrons. The correlated system is now described by the 
lattice model [Eq.(\ref{hamconduction})], so that we are allowed to study the distance $r$ 
dependence of correlation functions, 
like $\left\langle {\bf S}_f \cdot {\bf s}_r \right\rangle$.

Let us first derive the spin-spin correlations between the spin on the impurity site and on 
the central site $i=0$, i.e., $\left\langle {\bf S}_f \cdot {\bf s}_0 \right\rangle$. 
The DMRG results for various 
parameter sets are shown in Fig.\ref{fig3}(a) as function of the Coulomb interaction $U$. 
Since the antiferromagnetic correlation is derived from the {\it cf} exchange interaction, 
$\left\langle {\bf S}_f \cdot {\bf s}_0 \right\rangle$ is negative for all 
parameter sets and Coulomb interaction strengths. The absolute value of 
$\left\langle {\bf S}_f \cdot {\bf s}_0 \right\rangle$ increases with increasing 
$V$ and with decreasing $\left|\varepsilon_{\rm F}\right|$, as expected from the behavior of 
the binding energy. However, the influence of $\left|\varepsilon_{\rm F}\right|$ is rather 
smaller. In the limit $U \to 0$, $\left\langle {\bf S}_f \cdot {\bf s}_0 \right\rangle$ is 
antiferromagnetic but the magnitude is very small due to strong charge fluctuations, when 
the system is metallic [see inset of Fig.\ref{fig3}(a)]. It reflects the small binding energy 
around $U=0$. The magnitude of $\left\langle {\bf S}_f \cdot {\bf s}_0 \right\rangle$ 
increases with increasing $U$ and reaches its maximum value as $U \to \infty$, 
which means that one electron is localized on each site in that limit.

We consider next spin-spin correlations between a spin on the impurity site and on the 
next-nearest-neighbor site $i=1$, i.e., $\left\langle {\bf S}_f \cdot {\bf s}_1 \right\rangle$. 
In Fig.\ref{fig3}(b), the DMRG results for $\left\langle {\bf S}_f \cdot {\bf s}_1 \right\rangle$ 
are shown as a function of the Coulomb interaction 
$U$ for various parameter sets. One expects ferromagnetic correlations from the 
effective Hamiltonian (\ref{Heisham}) for finite values of $U$, 
and indeed $\left\langle {\bf S}_f \cdot {\bf s}_1 \right\rangle$ 
has positive sign for all the parameter sets and Coulomb interaction strengths. Note that 
the Ruderman-Kittel-Kasuya-Yosida (RKKY) interaction induces ferromagnetic correlations, 
as substitute for the spin-spin interaction (\ref{Heisham}), in the weak-coupling ($U \sim 0$) 
and metallic regimes. However, it is difficult to separate the contribution from RKKY and 
the interaction (\ref{Heisham}). The Coulomb interaction dependence of 
$\left\langle {\bf S}_f \cdot {\bf s}_1 \right\rangle$ is similar to that of 
$\left\langle {\bf S}_f \cdot {\bf s}_0 \right\rangle$. For the same parameter sets, 
the value of $\left\langle {\bf S}_f \cdot {\bf s}_1 \right\rangle$ is found to be slightly 
smaller than that of $\left|\left\langle {\bf S}_f \cdot {\bf s}_0 \right\rangle\right|$. 
This indicates a slow decay of the spin-spin correlation 
$\left\langle {\bf S}_f \cdot {\bf s}_r \right\rangle$ with distance $r$. It implies 
that the spin of the {\it f} electron is hardly screened by the spin on site $0$. In addition, 
the influence of $V$ on the spin-spin correlations is rather small. Note that the binding energy 
depends strongly on the hybridization $V$.

\begin{figure}[t]
    \includegraphics[width= 5.5cm]{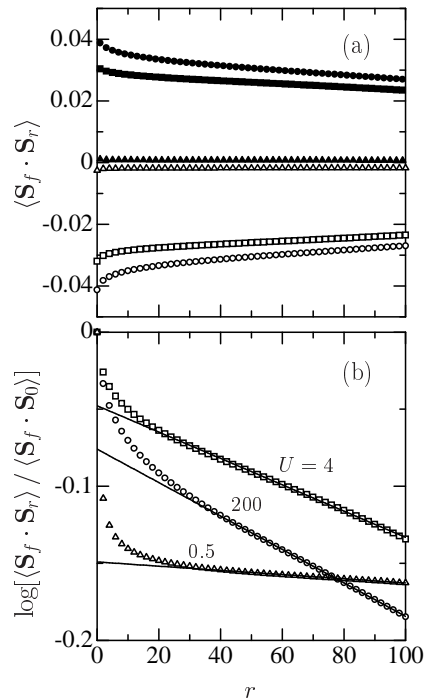}
  \caption{
(a) Spin-spin correlation functions $\left\langle {\bf S}_f \cdot {\bf s}_r \right\rangle$ 
as a function of the distance $r$ for $U=0.5$ (triangles), $4$ (squares), and $200$ (circles). 
(b) Semilogarithmic plot of the magnitude of the spin-spin correlation functions. 
The data is fitted by a function 
$\left\langle {\bf S}_f \cdot {\bf s}_r \right\rangle \simeq \exp(-\frac{r}{\xi})$ with 
$\xi = 3184$, $508$, $400$ for $U=0.5$, $10$, $200$, respectively.
  }
    \label{fig4}
\end{figure}

Let us now consider the distance dependence of the spin-spin correlation functions. 
In Fig.~\ref{fig4}(a), we plot the DMRG results for 
$\left\langle {\bf S}_f \cdot {\bf s}_r \right\rangle$ as a function of distance $r$ ($=|i|$). 
We choose three Coulomb interactions; (i) $U=0.5$ in the Kondo-singlet regime, 
(ii) $U=200$ in the limit of the local singlet regime, and (iii) $U=4$ in the intermediate 
regime where a maximal binding energy is obtained. The results for different distances are 
extrapolated to the thermodynamic limit $L \to \infty$ . We find that 
$\left\langle {\bf S}_f \cdot {\bf s}_r \right\rangle$ decays slowly and the sign 
changes alternately with $r$, i.e., $\left\langle {\bf S}_f \cdot {\bf s}_r \right\rangle$ 
has a positive (negative) sign for odd (even) $r$, denoted by solid (empty) symbols 
in Fig.~\ref{fig4}(a). The interaction (\ref{Heisham}) and/or the RKKY interactions cause 
ferromagnetic (antiferromagnetic) correlations between the spin of the {\it f} electron and that 
of the odd (even) site $r$. The absolute value of 
$\left\langle {\bf S}_f \cdot {\bf s}_i \right\rangle$ 
increases with increasing $U$  because larger Coulomb interactions stabilize the 
$2k_{\rm F}$-SDW oscillation which accompanies charge localization. 

Since the system is in a spin-gapped ground state, an exponential decay of the spin-spin 
correlation with distance must be expected. In Fig.~\ref{fig4}(b) we present a 
semilogarithmic plot of $\left\langle {\bf S}_f \cdot {\bf s}_r \right\rangle$ 
as a function of distance $r$. For a convenient comparison, we have normalized 
$\left\langle {\bf S}_f \cdot {\bf s}_r \right\rangle$ with respect to its value at $r=0$. 
The results can be fitted with a function $\exp(-\frac{r}{\xi})$ and thus the exponential 
decay of the correlation functions is confirmed for all values of $U$. The correlation lengths 
are estimated as $\xi = 3184$, $508$, $400$ for $U=0.5$, $4$, $200$, respectively. 
They seem to be much longer than those of other standard spin-gapped systems, e.g., 
$\xi = 3.19$ in the two-leg isotropic Heisenberg system. The large values of $\xi$ 
reflect exponentially small binding energies. They also mean that spin-polarized electrons are 
widely spread around the impurity site, i.e., the Kondo screening effect is quite weak. 
Furthermore, we note that the correlation functions decay rapidly around $r \simeq 0$. 
The decay rate is dependent of the magnitude of the {\it cf} exchange interaction.

\subsection{Less than half filling ($n<1$)}

We are also interested in doped systems, which are metallic even if $U>0$. 
We thus investigate the properties of the model~(\ref{hamiltonian}) with $\varepsilon_f=-3$ 
for various hole concentrations $n=1-N_{\rm h}/L$, where $N_{\rm h}$ is the number of 
doped holes ($N_{\rm h}>0$). 
For this choice of $\varepsilon_f$, the occupation number of the impurity site is near unity 
because the Fermi level lies well above $\varepsilon_f$. In the strong coupling 
limit ($U \gg t$), doubly occupied sites are excluded and therefore we can derive an 
effective model (\ref{hamiltonian}) by applying degenerate perturbation theory.~\cite{schork} 
The effective Hamiltonian is written as
\begin{eqnarray}
H = H_t + H_J + H_{\rm p} + H_{\rm K} + H^\prime.
\label{effhamiltonian}
\end{eqnarray}
Here $H_t$ is the kinetic-energy term of the conduction electrons 
\begin{eqnarray}
\nonumber
H_t &=& \sum_{i \sigma} t_i (\hat{c}^\dagger_{i+1 \sigma}\hat{c}_{i \sigma} 
+ \hat{c}^\dagger_{i \sigma}\hat{c}_{i+1 \sigma}), \\
t_i &=& -\frac{t}{2}\left(1-
\frac{V^2(2+2r+r^2)}{2\varepsilon_f^2(1+r)^2}\delta_{i0}\right),
\label{effkin}
\end{eqnarray}
with $\hat{c}^\dagger_{i \sigma}=c^\dagger_{i \sigma}(1-n_{i \sigma})$. 
Furthermore $H_J$ is a spin-coupling term between the conduction electrons, which is of 
the Heisenberg type
\begin{eqnarray}
\nonumber
H_J &=& J_i \sum_i {\bf s}_i \cdot {\bf s}_{i+1}, \\
J_i &=& \frac{2t^2}{U}\left(1-\frac{V^2}{\varepsilon_f(U-\varepsilon_f)}\delta_{i0}\right).
\label{effheis}
\end{eqnarray}
The sum of these two terms defines the 1D correlated electron system. It is essentially 
equivalent to a $t$$-$$J$ model except for small 
modifications around site $0$ due to the impurity. The term $H_{\rm p}$ corresponds to 
one-particle potential around the impurity site, which is given by
\begin{eqnarray}
\nonumber
H_{\rm p} &=& -\frac{\eta V^2}{2\varepsilon_f(1+r)}(1-n_0)
+ \frac{V^2t^2}{\varepsilon_f^2U(1+r)^2}\sum_{i=\pm 1}(1-n_i), \\
\eta &=& 2+r+\frac{2t^2}{\varepsilon_f^2(1+r)^2}(2+7r+7r^2+r^3).
\label{effpot}
\end{eqnarray}
It describes the attraction (repulsion) of a hole at site $0$($1$) by 
the {\it f} electron. Furthermore, $H_{\rm K}$ is a spin-spin interaction term in analogy to 
the {\it cf} exchange interaction,
\begin{equation}
H_{\rm K} = \frac{2\gamma V^2}{U-\varepsilon_f} {\bf S}_f \cdot {\bf s}_0 
+ \frac{tV^2(2+r)}{U\varepsilon_f(1+r)^2} 
{\bf S}_f \cdot \sum_{i=\pm 1}(\hat{\bf s}_{i0} + \hat{\bf s}_{0i}),
\label{effkondo}
\end{equation}
with $\hat{\bf s}_{ii^\prime}=(1/2)\sum_{\alpha\beta}
\hat{c}^\dagger_{i \alpha}{\bf \sigma}_{\alpha\beta}\hat{c}^\dagger_{i^\prime \beta}$, 
where ${\bf \sigma}_{\alpha\beta}$ are the Pauli matrices. 
The last term $H^\prime$ gives a correction to the effective model, 
\begin{equation}
\nonumber
H^\prime = \frac{2V^2t^2}{U\varepsilon_f^2(1+r)^2} 
\sum_{i=\pm 1} {\bf S}_f \cdot [{\bf s}_i(1-n_0) - {\bf s}_0(1-n_i)].
\label{effcorr}
\end{equation}
The first term of (\ref{effcorr}) implies an antiferromagnetic interaction between 
the impurity site and site $\pm1$ if there is a hole at site $0$; on the other hand, the second 
term gives a correction to the  Kondo-type interaction, i.e., the first term of 
(\ref{effkondo}), and the antiferromagnetic spin exchange between the impurity site and 
site $0$ may be reduced.

\subsubsection{Binding energy}

\begin{figure}[t]
    \includegraphics[width= 8.5cm]{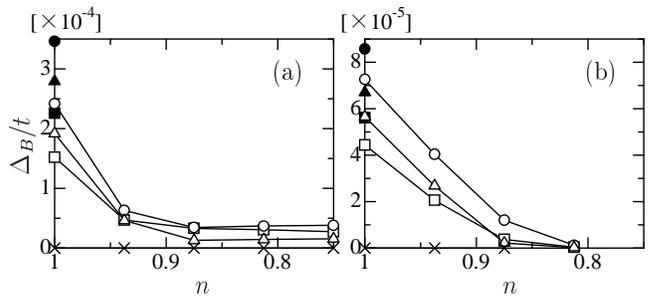}
  \caption{
Binding energy $\Delta_{\rm B}$ for (a) $V=0.2$ and (b) $V=0.1$ with $\varepsilon_f=-3$ 
as a function of the band filling $n$. The Coulomb interaction strengths are $U=0$ (crosses), 
$2$ (triangles), $4$ (circles), and $10$ (squares). Filled (empty) symbols correspond to 
the data for $n=1$ ($n<1$) and empty symbols at $n=1$ represent the values for infinitesimally 
doped systems.
  }
    \label{fig5}
\end{figure}

Of particular interest is the evolution of the binding energy of the impurity-induced bound 
state upon hole doping. We can easily imagine that the binding energy is suppressed by hole 
doping due to the enhancement of charge fluctuation. Thus, away from half filling, the 1D 
correlated system is metallic and the bound state changes from a local singlet 
to the Kondo singlet. If the bound state survives with hole doping, it has a 
much larger energy than the standard Kondo singlet.
In Fig.~\ref{fig5} we show the binding energy $\Delta_{\rm B}$ as a function of band 
filling $n$ ($ \le 1$) at (a) $V=0.2$ and (b) $V=0.1$ with $\varepsilon_f=-3$ for various 
values of $U$. Filled (empty) symbols refer to the data for $n=1$ 
($n<1$) and empty symbols at $n=1$ represent the values for infinitesimally doped systems 
(see below). Roughly speaking, $\Delta_{\rm B}$ is discontinuously reduced at $n=1$ and 
decreases with increasing hole doping for all cases except $U=0$. We find however that 
$\Delta_{\rm B}$ remains of the same order of magnitude as in the half-filled 
case even at doping level up to a few percents. Also, the dependence of $\Delta_{\rm B}$ on $U$ 
is weaker for higher doping concentrations.

More precisely, there are two differences in behavior on the hybridization strength $V$. 
One is that in the vicinity of $n=1$ the binding energy for $V=0.2$ decreases 
more rapidly than that for $V=0.1$ despite larger {\it cf} exchange coupling 
[Eq.(\ref{effkondo})]. It must be associated with the attraction between doped holes and 
the {\it f} electron, which is described in detail in the next paragraph. The other is that 
the binding energy disappears at lower doping levels for small values of $V$; $\Delta_{\rm B}$ 
for $V=0.2$ maintains its value at $n \lesssim 0.9$ and that for $V=0.1$ goes to zero around 
$n \simeq 0.8-0.9$. It results from the size of the {\it cf} exchange coupling $J_{\it cf}$, 
and thus the critical doping concentration is highest at $U \approx 4$ giving a maximal value 
of $J_{\it cf}$.

For the limit $n \to 1$ we have extrapolated the finite-size binding energy $\Delta_{\rm B}(L)$ 
to the thermodynamic limit $L \to \infty$ for the four-hole doped system by going up to $L+1=510$. 
One notices that the value of the binding energy in the limit $n \to 1$ differs from the $n=1$ 
undoped value. It reflects the fact that the binding energy of the Kondo singlet in the 
infinitesimally doped system is less than that of the local singlet in the undoped system. 
The reason being that, when the system is doped by a hole, the career tends to move onto 
site $0$ due to the attraction 
from the impurity site [Eq.(\ref{effpot})] and thus a spin-singlet formation is prevented. 
The discontinuity is higher for $V=0.2$ than for $V=0.1$ because the attractive 
interaction is enhanced by the hybridization $V$. Such a discontinuity of the spin-excitation 
energy has also been found in studies of ladder systems.~\cite{poilblanc2,nishimoto}
Note that in the hole-doped case the $V$ dependence of the binding energy is not simple 
because $V$ enhances two competing effects: (i) the attraction between doped holes and the 
{\it f} electron and (ii) the {\it cf} exchange coupling between conduction electrons and 
the {\it f} electron.

\subsubsection{Spin-spin correlations}

\begin{figure}[t]
    \includegraphics[width= 8.5cm]{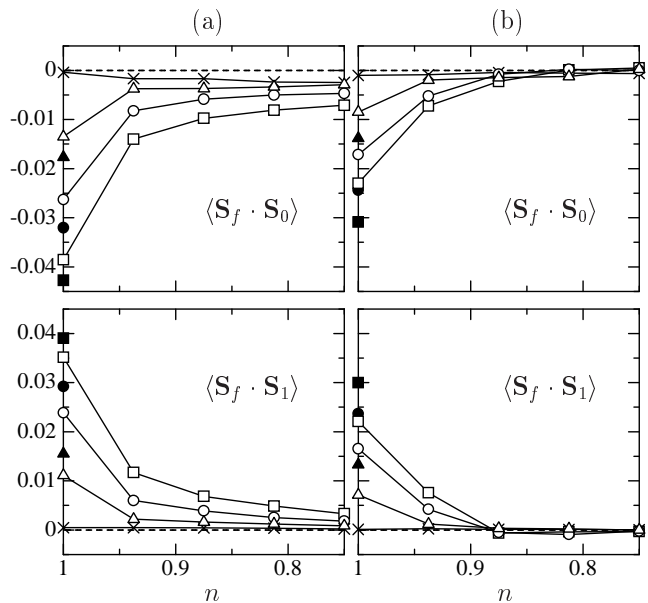}
  \caption{
Spin-spin correlation functions $\left\langle {\bf S}_f \cdot {\bf s}_0 \right\rangle$ and 
$\left\langle {\bf S}_f \cdot {\bf s}_1 \right\rangle$ for (a) $V=0.2$ and (b) $V=0.1$ with 
$\varepsilon_f=-3$ as a function of the band filling $n$. The Coulomb interaction strengths 
are $U=0$ (crosses), $2$ (squares), $4$ (triangles), and $10$ (circles). Filled (empty) 
symbols correspond to the data for $n=1$ ($n<1$) and empty symbols at $n=1$ represent the values 
for infinitesimally doped systems.
   }
    \label{fig6}
\end{figure}

Finally, we study the hole-doping dependence of spin-spin correlations between the 
{\it f}- and conduction electrons. The correlation is expected to be weakened by 
hole doping due to an increase of charge fluctuations. In Fig.~\ref{fig6}, we show the 
spin-spin correlation functions $\left\langle {\bf S}_f \cdot {\bf s}_0 \right\rangle$ and 
$\left\langle {\bf S}_f \cdot {\bf s}_1 \right\rangle$ as a function of band filling 
$n$ ($ \le 1$) when (a) $V=0.2$ and (b) $V=0.1$ with $\varepsilon_f=-3$ for various Coulomb 
interaction strengths. The properties are fundamentally linked to those of the binding energy 
as follows: (i) correlations are suppressed by hole doping and (ii) there exists a discontinuity 
at $n=1$. 

Let us now investigate the DMRG results for the two $V$ values. When $V=0.2$, all the correlation 
functions for finite $U$ decrease rapidly close to $n=1$ and decay slowly when $n \lesssim 0.9$. 
This behavior is quite similar to that of the binding energy. It is seen that 
$\left|\left\langle {\bf S}_f \cdot {\bf s}_1 \right\rangle/\left\langle {\bf S}_f \cdot {\bf s}_0 \right\rangle\right|$ decreases with decreasing $n$. The small value corresponds to a rapid 
decay of $\left\langle {\bf S}_f \cdot {\bf s}_r \right\rangle$ around $r=0$, as seen in 
Fig.~\ref{fig4}, e.g., for $U=0.5$ and $n=1$. It is accompanied by a transfer from the local singlet 
to the Kondo singlet. It also suggests a 
reduction of the RKKY interaction with doping. In addition, it is surprising that 
$\left\langle {\bf S}_f \cdot {\bf s}_0 \right\rangle$ seems to be enlarged by hole doping for 
small values of $U$ ($\lesssim 2$). The ``exchange hole'' around the impurity is as a 
consequence of the Pauli principle. When $V=0.1$, all the correlation functions decrease 
monotonously and go to zero around $n \simeq 0.9$, which is accompanied by a vanishing of 
the binding energy. For $n \lesssim 0.8-0.9$, 
$\left|\left\langle {\bf S}_f \cdot {\bf s}_1 \right\rangle\right|$ has small 
negative values for large values of $U$, which indicates antiferromagnetic correlations 
between the {\it f} electron 
and the spin at site $1$. It is derived from the first term of (\ref{effcorr}) and 
was previously suggested in Ref.5.%\cite{schork}.

\section{CONCLUSION}

Using the DMRG method, we have studied a magnetic impurity embedded in a correlated electron system 
which is assumed to be the 1D Hubbard chain. At half filling, we confirm that the binding 
energy increases exponentially in the weak-coupling regime. There is a crossover from 
the Kondo singlet to the local singlet. The former state involves a wider spread of spin-polarized 
electrons around the impurity than the latter one. 
With increasing values of $U$, the binding energy has a maximum around $U \approx 4$ 
and afterwards decreases inversely proportional to the Coulomb interaction. 
Due to the formation of a singlet bound state, the spin-spin 
correlation function decays exponentially with distance from the impurity site for all values of 
$U$ ($>0$). The correlation length is quite long when the binding energy is small. 
It becomes shorter with increasing Coulomb interaction. For 
infinitesimally hole doping, we find a discontinuous reduction of the binding energy and of the 
spin-spin correlations from the values at half filling. For further doping, the binding energy 
is reduced but remains of the same order of magnitude as in the half-filled case even for doping 
concentration of a few percent.  The electron-doped case is not studied here, but we expect 
qualitatively similar properties as for hole doping. However, there is no discontinuity at half filling. This is so because when an electron is added to the half-filled system it is distributed uniformly over the 1D chain 
except for site $0$.

\acknowledgments
We thank T.~Takimoto for useful discussions.

\end{document}